# Galaxies with Abnormally High Gas Content in the Disk

A. V. Zasov[a, b, *] and N. A. Zaitseva[b]

[a] *Faculty of Physics, Lomonosov Moscow State University, Moscow, 119991 Russia*
[b] *Sternberg Astronomical Institute, Lomonosov Moscow State University, Moscow, 119234 Russia*
*\*e-mail: zasov@sai.msu.ru*

**Abstract**—The content of gas in galaxies with an anomalously high relative mass of hydrogen $M_{\rm HI}/M_*$ for a given mass of the stellar population $M_*$ (VHR-galaxies) is considered, using the available samples of such galaxies. It is shown that, within the optical diameter $D_{25}$, the mass of HI in VHR galaxies, as well as in galaxies with "normal" HI content, is limited by a value that depends on the specific angular momentum of the disk. Outer gaseous disks beyond $D_{25}$, which contain the main amount of HI in most of the galaxies we consider, are gravitationally stable, and, as a rule, they retain an approximately constant value of the stability parameter $Q_{\rm gas}$ over a large range of radial distances. It allows to propose that the outer disks of VHR galaxies are not recently acquired, but are of great age, and their gravitational instability was the main regulator of star formation during their formation. In this case, the extended disks of galaxies should also include a low-brightness stellar components of old stars extending far beyond their optical diameter $D_{25}$.

## 1. INTRODUCTION

The content of gas in the disks of galaxies is a key factor determining the current activity of star formation and, consequently, the nature of the evolution of galaxies. The main amount of gas in galactic disks in the present era belongs to atomic hydrogen HI, except for the inner regions of galaxies, where molecular gas often predominates (see, for example, [1]).

The radial distribution of gas density in galactic disks, like their mass, depends on a number of external and internal factors, and must have changed in the course of evolution. A mass of gas within a decreases as a result of star formation, as well as due to activity of massive stars, calling the ejection of gas from a disk. On the other hand, the accretion of gas onto a galaxy, the absorption of satellites containing gas, as well as the fall of the cooling halo gas onto a disk, in principle, can compensate for the loss of gas, maintaining its content at a certain level, where these processes are approximately balanced (the so-called "bathtub model", see, for example, [2]). Obviously, these factors play different roles depending on the mass of a galaxy, its structure and kinematics, the rate of star formation, and also on its immediate environment. It would seem that this should have led to a wide variety of the observed values of the relative mass of gas and its radial distribution in galaxies. However, paradoxically, despite the variety of conditions under which the evolution of the gas composition occurs in different galaxies, their HI mass (at least in galaxies with ongoing star formation) turns out to be a quite predictable value, since it correlates with such conservative internal parameters as the diameter of the optical disk ($D_{25}$) associated with the B-isophote of 25 st.magn./sq.s, the rotation speed $V_{\rm rot}$, or the specific angular momentum of the disk, proportional to $V_{\rm rot}D_{25}$ (see discussions of the issue in [3] and references therein). At the same time, the radial distribution of surface density of HI (if we exclude the central region of the disk, where molecular gas often predominates) in most galaxies has a similar profile, despite the difference in galaxies in size or mass, if we normalize the radial coordinate to the optical radius or the radius of the gaseous disk $R_{\rm HI}$ [4, 5].

The relative mass of neutral hydrogen $\mu = M_{\rm HI}/M_*$ in spiral galaxies also correlates with the stellar population mass $M_*$, systematically decreasing with growth of $M_*$ (see, for example, [5–7]). The relationship between $\mu$ and $M_*$ along with the dependence between $M_{\rm HI}$ and the luminosity of galaxies is taken as a reference for identifying galaxies with anomalously high gas content. Below, for brevity, we will call these galaxies VHR (very HI-rich) galaxies.

Samples of galaxies with an anomalously high HI mass are presented in several papers. In [8], within the framework of the Bluedisk project, a sample of 25 massive galaxies ($\log M_*/M_\odot > 10$) with active star formation, which have an anomalously high value of the relative gas mass $\mu$ for a given stellar population mass, was studied on the WSRT radio interferometer. It was shown that these galaxies have extended HI disks, which, as a rule, contain the bulk of the gas mass, the diameters $D_{HI}$ of which (within the accepted threshold density 1 $M_\odot/pc^2$) are several times greater than the optical diameters $D_{25}$. The most extended HI disks reach a diameter of $\approx 100$ kpc and $M_{HI} \approx 2 \times 10^{10} M_\odot$. Nevertheless, the VHR galaxies of this sample lie on the continuation of the dependence $M_{HI}-D_{HI}$ for normal galaxies. Note that, according to the selection conditions, the Bluedisk sample is biased towards objects with intense star formation.

A number of high-luminosity VHR galaxies ($M_K < -22$) were identified and studied by the authors in [9–11]. These galaxies were selected on the basis of the Parkes HI blind survey (HI eXtreme (HIX) galaxy survey). According to the accepted selection criterion, the mass $M_{HI}$ in them is at least 2.5 times higher than the value expected from the dependence $M_{HI}-L_R$ (where $L_R$ is the luminosity in the $R$-band) for the control sample. The dependence $M_{HI}-D_{HI}$ for these galaxies was extended even further than in [8] (up to $M_{HI} \approx 10^{11} M_\odot$), although in the region of large diameters it is characterized by a higher dispersion of values. Their optical sizes, when compared with the control sample, turned out to be normal for their luminosity. The latter indicates that the galaxies with a large mass of HI considered here have normal stellar disks.

In [7], the authors identified 34 VHR galaxies (HIghMass sample) with a small redshift from the radio survey of the sky in the HI line on the Arecibo radio telescope (ALFALFA survey). The selected galaxies have a large mass of HI ($M_{HI} > 10^{10} M_\odot$) and an anomalously high value of $\mu$ for a given mass of their stellar population. The optical properties of these galaxies were studied and it was shown that in most cases they are characterized by active star formation in the stellar disk.

Another sample of VHR galaxies was studied in [12]. There, from the GASS survey [6], a population of relatively close galaxies with $0.025 < z < 0.05$ was selected, having both a high mass of hydrogen ($\log M_{HI}/M_\odot > 10$) and a high mass of the stellar population, $\log M_*/M_\odot > 10.5$, but at the same time the sample galaxies are characterized by a low specific rate of star formation sSFE = SFR/$M_*$. The obtained HI distribution maps are consistent with the conclusion that the low sSFE value for the selected galaxies is due to the low surface density of their extended gaseous disks, where there is no active star formation, while in the inner regions of the disks star formation appears to occur in the same way as for ordinary spiral galaxies.

A number of studies have considered various options that could potentially explain the abnormally high gas content in the disks of massive galaxies, located, as a rule, far beyond their optical radius: it may be a merging with gas-rich galaxies (small merging), an accretion of gas onto a disk from gas filaments, a gas falling from the halo as a result of its cooling, or low rates of gas consumption for star formation due to its low density associated with a large specific angular momentum of the gas disk (see discussion of the issue in [8, 10–14]). In [10], the authors used the disk-averaged value of the stability parameter for galaxies from the NIX survey, and concluded that the large mass of gas in the disk is associated with a high specific angular momentum of the gas layer, which ensures its gravitational stability.

It remains an open question whether the outer gaseous disk was acquired already after the formation of the stellar component of the galaxy, or whether it arose together with the galaxy from gas posessing a very high angular momentum and has been preserved to the present due to the low efficiency of star formation. Apparently, different scenarios are possible in different galaxies.

Note that the excess amount of gas in VHR galaxies concerns only atomic hydrogen. The mass of molecular gas in them, where it was found, turned out to be several times less than the mass of HI [15, 16]. As in ordinary spiral galaxies, this gas is concentrated in the inner region of the optical disk.

In this paper, we analyze in more detail the gas content in VHR galaxies both within the optical boundaries of stellar disks (Section 2) and in the extended outer gaseous disks of these galaxies (Section 3), as well as the relationship between the surface gas density and the condition of the gravitational stability of the disk. Section 4 includes a discussion and summary.

The parameters depending on the accepted distance are given to the distance scale corresponding to the parameter $H_0 = 75$ km/s/Mpc.

## 2. THE MASS OF HI WITHIN THE OPTICAL RADIUS

Let us consider in more detail what limits the mass of gas contained in the disks of galaxies. The dependences between the mass of gas and the dynamic characteristics of galaxies mentioned in the Introduction can be explained, at least on a qualitative level, if we assume that the azimuthally averaged values of the surface density of gas in the disk in star-forming galax-ies are related to the angular momentum of the rotating gas layer and the density of the stellar-gas disk at a given distance $R$ from the center, since these parameters determine the condition for the development of the gravitational stability of the disk on kiloparsec scales (see the discussion of the issue, for example, in [3, 10, 17–19]). Analysis of the local gravitational instability of an inhomogeneously rotating thin gravitating disk, first carried out in [20] for a protoplanetary Keplerian disk, as applied to a collisionless (stellar) disk [21] and to a gaseous disk [22], leads (under some simplifying assumptions) to an approximate relation that should be fulfilled for radial perturbations at a marginal disk stability:

$$\Sigma_{\text{gas}}^c = \frac{c_g \varkappa}{Q_{\text{gas}}^c \pi G}, \qquad (1)$$

where $c_g$ is the one-dimensional dispersion of turbulent gas velocities, $\varkappa(R) \approx \sqrt{2}V/R$ is the epicyclic frequency under the assumption of a "flat" rotation curve, and $Q_{\text{gas}}^c$ is the Toomre parameter for marginally stable gaseous disk.[1] Its value is equal to unity for radial perturbations of a thin disk, and, as theoretical and model estimates show, it reaches 1.4–2 (depending on the distance from the center) when nonradial perturbations are taken into account (see, e.g., [3, 23] and references in these works).

When passing from the density of an atomic gas to the total density of the gas, it must be considered that some part of its mass belongs to elements heavier than hydrogen, which increases the estimate of the mass of the gas by about 35%, and a part of the gas is in the form of molecular hydrogen, which is much more difficult to take into account. The few available estimates of the abundance of $H_2$ in VHR galaxies lead to values of $M_{H_2}/M_{HI}$ from a few percent to 30% [15, 16, 24]. As in the case of late-type normal spiral galaxies, where the molecular gas mass is, as a rule, comparable to the mass of HI, in VHR galaxies the molecular gas is also concentrated predominantly in the inner regions. Considering molecular hydrogen and heavy elements, we assume that inside the optical disk of galaxies the coefficient $\eta = M_{HI}/M_{\text{gas}} \approx 0.5$, that is, HI accounts for about half of the cold gas. For the extended outer HI disks, this value can be considered as the lower limit (the upper limit corresponding to the absence of molecular gas is approximately equal to 0.7).

Observations show that the dispersion of turbulent gas velocities in galactic disks of different masses varies little from galaxy to galaxy. It does not explicitly depend on the gas content in the disk and remains constant or slowly decreases with distance from the center of the galaxy, amounting to $c_g \approx 10$ km/s in the outer regions of the stellar disks (see the discussion in [1, 25, 26]).

Since the velocity of rotation of the gas on most of the disk, as a rule, varies little with distance from the center, then, as it was shown earlier, at a gas density equal to (or proportional to) the critical density defined by equation (1), the gas mass within the fixed (for example, optical) radius should be proportional to the product of the rotation velocity and the value of this radius (see [3] and references therein to earlier publications). This conclusion agrees well with observations of galaxies of late morphological types (as it was demonstrated for galaxies in the Local Volume in [27]). In a more general case, given that the gas density may be below the critical value, the upper limit of the HI mass will be

$$\begin{aligned} M_{HI}(D_{25}) &\leq \int_0^{D_{25}/2} 2\pi\eta R \Sigma_g(R) dR \\ &= \sqrt{2}\pi\eta \frac{c_g}{Q_{\text{gas}}^c G} V_{\text{rot}} D_{25}, \end{aligned} \qquad (2)$$

where the coefficient $\eta = M_{HI}/M_{\text{gas}}$ considers the contribution of helium and molecular hydrogen to the total mass of the gas (see above). Taking $c_g \leq 10$ km/s, and the Toomre parameter $Q_{\text{gas}}^c \geq 1$, for the mass of HI we obtain the relation

$$\log M_{HI}(D_{25}) \leq 6.21 + \log(V_{\text{rot}} D_{25}), \qquad (3)$$

where the velocity $V_{\text{rot}}$ is expressed in km/s, the diameter $D_{25}$ is in kpc, and $M_{HI}$ is in solar masses.

For comparison with observational data, we chose VHR galaxies for which there are data in the literature on the distribution of surface densities of neutral hydrogen and rotation velocities (found from the width of the HI line or from the rotation curve), as well as the estimates of the integral magnitude in the $K$ filter [28]. The main parameters of galaxies are presented in Table 1.

In Fig. 1, the mass of HI in the optical disks of galaxies is compared with the product $V_{\text{rot}} D_{25}$ characterizing the specific moment of rotation for VHR galaxies (the data were taken from [7, 10, 15, 24]) and "normal" galaxies of the THINGS sample [1]. The masses $M_{HI}$ of the compared samples belong to the region $R_{25} = D_{25}/2$.

As can be seen from Fig. 1, the expected value of the upper limit of the mass of HI, obtained from the condition of the gravitational stability of the gas layer within the optical radius, really limits the distribution of points in the diagram, only a few of which are above the straight line corresponding to equation (3). At the same time, VHR galaxies (red symbols) are not distinguished in the diagram among normal galaxies by higher values of the gas mass. Almost all galaxies are located below the line, which indicates the stability of

---

[1] For a collisionless (stellar) disk, instead of the number $\pi$, there should be 3.36, but this difference is insignificant.

758 ZASOV, ZAITSEVA

**Table 1.** Sample of VHR galaxies

| PGC | Other designation | $D$, Mpc | $\log M_{\rm HI}^{\rm full}$, $M_\odot$ | $\log M_{\rm HI}(R_{25})$, $M_\odot$ | $\log M_{\rm HI}/M_*$ | $R_{\rm HI}/R_{25}$ |
|---|---|---|---|---|---|---|
| 624 | ESO111–G014 | 103.8 | 10.6 | 9.86 | 0.093 | 2.2 |
| 2887 | ESO243–G0021 | 118.5 | 10.7 | 9.72 | −0.014 | 3.4 |
| 3089 | NGC 289 | 21.7 | 10.5 | 9.3 | −0.049 | 7.0 |
| 7298 | ESO245–G010 | 76.7 | 10.5 | 9.89 | −0.075 | 2.2 |
| 11691 | ESO417–G018 | 63.3 | 10.4 | 9.92 | 0.14 | 2.3 |
| 14631 | ESO055–G013 | 98.5 | 10.3 | 8.92 | 0.25 | 5.0 |
| 21343 | ESO208–G026 | 39.8 | 9.9 | 8.64 | 0.09 | 4.4 |
| 35288 | ESO378–G003 | 40.3 | 10.3 | 9.06 | 0.2 | 4.6 |
| 42463 | ESO381–G005 | 75.9 | 10.4 | 9.24 | 0.21 | 3.3 |
| 63796 | ESO461–G010 | 89.5 | 10.3 | 9.11 | 0.4 | 1.4 |
| 66678 | ESO075–G006 | 141.4 | 10.9 | 9.72 | 0.29 | 4.4 |
| 70281 | ESO290–G035 | 78.5 | 10.5 | 9.45 | −0.11 | 1.6 |
| 33645 | UGC 6168 | 107.6 | 10.35 | 9.95 | −0.02 | 2.1 |
| 42836 | UGC 7899 | 115.9 | 10.42 | 10.18 | −0.07 | 1.8 |
| 50455 | UGC 9037 | 79.2 | 10.33 | 9.88 | 0.24 | 2.6 |
| 71078 | UGC 12506 | 96.5 | 10.53 | 10.05 | 0.07 | 2.5 |

$D$ is the accepted distance to the galaxy; $\log M_{\rm HI}^{\rm full}$ is the total HI mass in the galaxy according to [7, 9]; $\log M_{\rm HI}(R_{25})$ is the HI mass inside the optical radius $D_{25}/2$; $\log M_{\rm HI}/M_*$ is the ratio of the total HI mass to the stellar mass according to [7, 9]; $R_{\rm HI}/R_{25}$ is the ratio of the HI radius to the optical radius.

their gaseous disks: the parameter $Q_{\rm gas}$ for them lies in the range 1–5, assuming its constancy along the radius. Note that this applies only to the current state of galaxies. The characteristic gas exhaustion time is several billion years; therefore, in the past, disks could contain several times more gas and be in the region of critical values for the gas mass.

In all the galaxies considered here, the active star formation is observed. Therefore, those that are located in the diagram near the upper limit for the mass of gas can be considered as candidates for galaxies where a high content of gas is (or was) supported by accretion. For the galaxy to remain near the upper boundary of $M_{\rm HI}$, the gas acquisition rate must be no less than the gas consumption for star formation (several solar masses per year). Of the considered VHR galaxies, this may be primarily applied to the galaxies of the HIghMassSample sample (marked with red stars in the diagram), for which the star formation rates were estimated in [7].

Relation (3) allows us to pass to the dependence of the relative mass of HI on the mass of the stellar population $\mu$–$M_*$, which is important for the selection of VHR galaxies. To do this, we will use the observed close correlation between the specific angular momentum and the IR luminosity (and, hence, the mass $M_*$) of the stellar population of galaxies (see Fig. 2). The regression line $Y(X)$ may be written as:

$$\log L_K = 1.39\log(V_{\rm rot} D_{25}) + 5.60. \quad (4)$$

Combining this expression with the above dependency (3), we get:

$$\log(M_{\rm HI}(D_{25})/L_K) \leq 2.18 - 0.28\log L_K. \quad (5)$$

Figure 3 shows the diagram $\log \dfrac{M_{\rm HI}(D_{25})}{M_*}$–$\log M_*$ inside $D_{25}$ for "normal" galaxies of the THINGS sample (yellow circles) and VHR galaxies (red circles), where the ratio $M_*/L_K = 0.5$ for the stellar population of galaxies with a normal abundance of heavy elements was taken (following [1]) to pass from $L_K$ to $M_*$ (see also [30]). For comparison, black dots show the positions of isolated late-type galaxies (AMIGA sample). As follows from the diagram, the assumption that the surface density of gas is proportional to the Toomre critical density is consistent with a decrease in the gas mass in galaxies along the sequence of their stellar masses, and the upper limit of the gas mass in galaxies of the types under consideration is consistent with what is expected for a marginally stable gas layer.

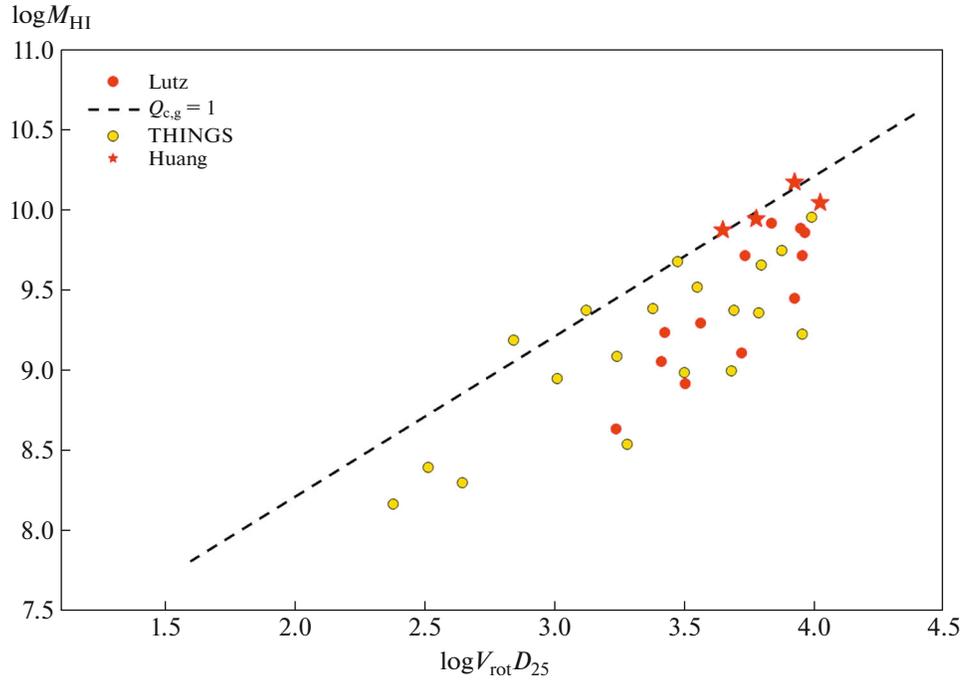

**Fig. 1.** Relationship between the hydrogen mass $M_{HI}$ inside the optical diameter $D_{25}$ and the product $V_{rot} D_{25}$, which characterizes the specific angular momentum of the disk, for VHR galaxies (red symbols) and galaxies of the comparison sample THINGS (yellow symbols). Red circles are galaxies from surveys [9–11], red stars are galaxies from the sample of Huang et al. [7], considered in [15, 24]. The dashed line reflects the expected ratio for marginally stable disks according to (2).

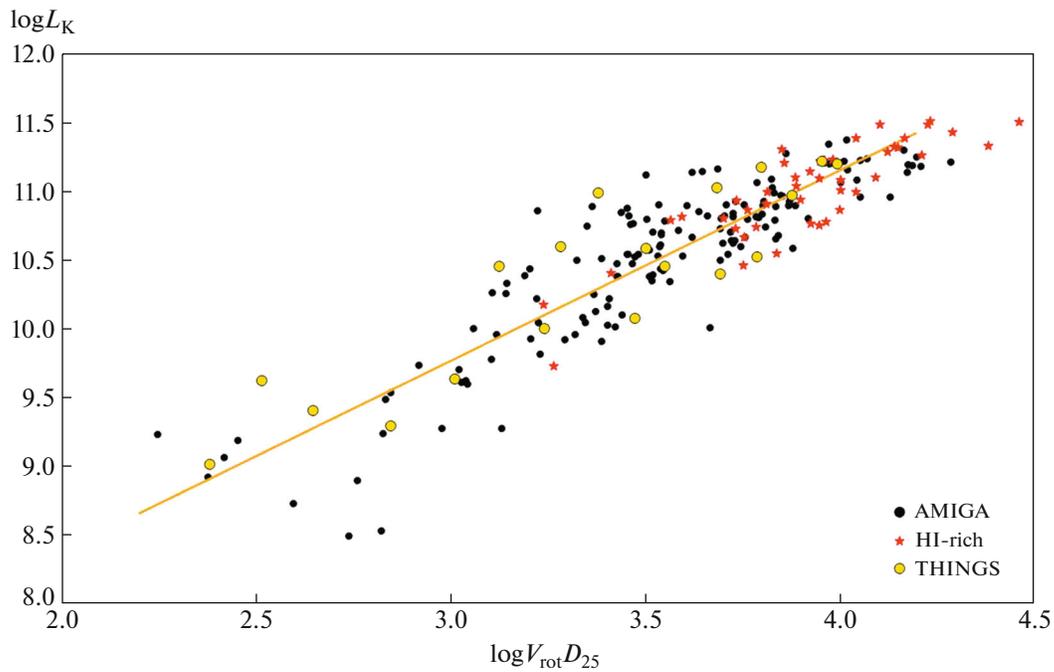

**Fig. 2.** Relationship between the luminosity $L_K$ in the $K$ filter and specific angular momentum $V_{rot} D_{25}$. Black dots are isolated galaxies from the AMIGA sample [29], red dots are a sample of VHR galaxies considered in this work, and yellow circles are the comparison sample THINGS [1]. The yellow line is the regressive for isolated galaxies.

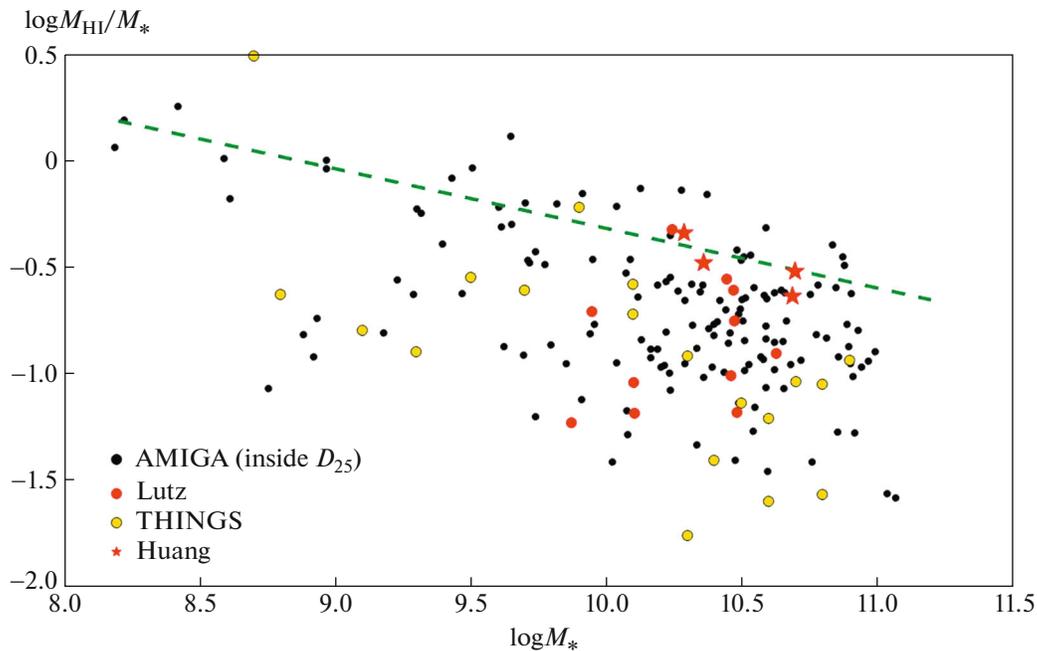

**Fig. 3.** Relationship between the ratio of the HI mass inside the optical radius to the stellar mass and the mass of stellar population. Black dots are isolated galaxies from the AMIGA sample [29], red circles mark the sample of VHR galaxies, and yellow circles are the comparison sample THINGS [1]. The green dashed line corresponds to the ratio between the parameters for a marginally stable disk.

## 3. OUTER GASEOUS DISKS OF VHR GALAXIES

Let us take a closer look at the outer gaseous disks of VHR galaxies. Their gas density $\Sigma_{HI}(R)$ slowly decreases with distance from the center, so that HI is often traced to distances many times exceeding the optical radius $R_{25}$ (see Table 1). At the same time, if the radial distribution of gas density within the limits of $R_{25}$ in most galaxies has a similar character, then the gas content outside this radius is very diverse. In the absence of accretion or external influence on the gaseous component of a galaxy, the gas density can remain approximately at the same level for billions of years as a result of the low efficiency of star formation in a very rarefied medium at a large distance from the center of a galaxy. Therefore, the outer gaseous disks can be relatively young as well as formed together with a galaxy.

Some VHR galaxies exhibit a weak optical continuation of a low-brightness disk far beyond $R_{25}$ in the visible and *UV* regions of the spectrum, while HI is traced even further from the center. This outer optical disk may be structureless, but in some cases the extended low-contrast spirals are observed up to a distance of $(1.5–3)R_{25}$ from the center (the examples: PGC 66678, PGC 42463, PGC 35288, PGC 7298, PGC 2887).

Figure 4 shows the radial surface density profiles $\Sigma_{HI}(R)$ for galaxies in the HIX survey [10] and for four galaxies from the HIghMassSample according to [15, 24] (PGC 50455, PGC 42836, PGC 33645, and PGC 71078). The radial coordinate is normalized to the optical radius $R_{25}$. For comparison with "normal" galaxies, the dark circles show the radial profile of the HI density averaged over a large number of nearby late-type galaxies in [5] and recalculated to the coordinate $R/R_{25}$. Here we used the average value $R_{HI}/R_{25} = 1.35$ for a sample of nearby spiral galaxies whose disks are observed almost "flat" [31].

As can be seen from Fig. 4, the density $\Sigma_{HI}$ decreases with distance from the center of VHR galaxies more slowly than for ordinary spirals. Nevertheless, within the optical disk ($R/R_{25} = 1$) in most cases, it lies in the interval $(4–10)\,M_{\odot}/pc^2$, which is typical for spiral galaxies. Two galaxies, PGC 70281 and PGC 63796, are distinguished by the lowest gas densities in the region $R/R_{25} \leq 1$. These galaxies are oriented "edge-on" to us, so their gas mass is apparently underestimated due to the low transparency of the outer gaseous disks in the HI line along the line of sight in the plane of the extended disk. In the HyperLeda database [32], the self-absorption correction for edge-on galaxies was assumed to be $\delta M_{HI} = -0.82$ (in stellar magnitudes) [33], which corresponds to approximately a twofold attenuation of the flux (in [3], a slightly lower value was obtained for the attenuation of HI). However, in the case of the extended hydrogen disks containing the bulk of the gas, the self-absorption in the HI line when the disk is edge-on should be much higher than for ordinary spirals.

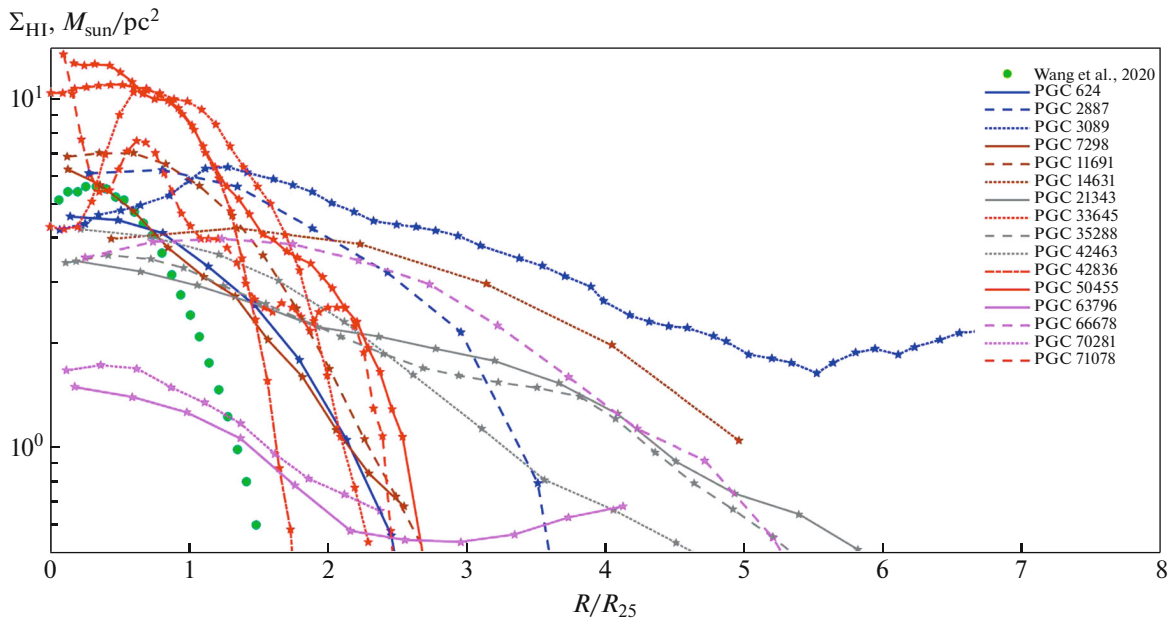

**Fig. 4.** The HI surface density profiles for the VHR galaxies under consideration in comparison with the averaged profile for close galaxies of late morphological types from [5] (green circles) for the accepted value $R_{\rm HI}/R_{25} = 1.35$. The distance from the center is expressed in units of optical radius $R_{25}$.

Note that among the galaxies under consideration there is another one oriented almost "edge-on"—PGC 71078 (=UGC 12056), but it nevertheless demonstrates a relatively high gas density. However, this is the only galaxy under consideration where the density of HI drops rapidly even within the optical disk, so that its outer gaseous disk contains a relatively small mass of HI.

## 4. DISCUSSION AND CONCLUSIONS

As noted earlier in [10] (where the HIX sample galaxies was taken as an example), galaxies with anomalously high HI abundances do not noticeably stand out in terms of morphology, star formation rates, or surface brightness (density) of their stellar disks from "normal" galaxies with a similar disk mass. A similar conclusion was made for the VHR galaxies of the Bluedisk sample [8]: the average surface brightness of galaxies within the optical diameter, as well as the star formation rate per unit area of the stellar disk, have similar intervals of values in the VHR galaxies and in the control sample galaxies. The star formation efficiency (the star formation rate per unit mass of the HI + $H_2$ gas) within optical disks in ordinary spiral galaxies and in VHR galaxies also behaves in a similar way, decreasing towards the periphery of optical disks

(the HIghMass sample [7]). The metallicity of gas in VHR galaxies, where it was possible to directly measure it, also turned out to be common for spiral galaxies, if to compare disk regions with similar stellar population and HI column densities [11]. This makes intense accretion of gas within the optical disk unlikely, although there may be exceptions for some galaxies where the HI mass is anomalously high within the optical radius (see comments to Fig. 1 in the text).

The present work also shows that within the optical radius the considered VHR galaxies, these are systems that (with some exceptions) do not differ either in optical characteristics or in the relative gas content from ordinary spiral galaxies. The main feature of most VHR galaxies is the presence of the extended massive gaseous disks.[2]

A high gas content of VHR galaxies makes their outer regions similar to the disks of giant low-brightness spiral galaxies (GLSB galaxies), which also have surface densities of HI about a few tenths of $M_\odot/\rm pc^2$ and a weak star formation. The local values of the HI velocity dispersion in the extended gaseous disks of these galaxies are also close to 10 km/s, although they are comparable with the spectral resolution of radio observations, which does not allow a more accurate estimate [35]. There are apparently no fundamental differences between the extended gaseous disks of low- and high-brightness galaxies. Note that some GLSB galaxies may also contain a small inner stellar disk with a surface brightness typical for ordinary galaxies. An example is the GLSB galaxy UGC 1378, whose spectral and photometric studies are given in [36]. In all respects, it can be attributed to VHR galaxies. Several GLSB galaxies with an extended optical (and gaseous) disks are included in the HIghMass sample studied in [15, 24]. A striking example is the galaxy AGC 19204, which, due to its very high gas content and the presence of the extended outer disk, can be attributed to objects of the Malin-1 type [13].

---

[2] The extended outer disks are observed not only in galaxies with an anomalously high integral mass of HI, but also in some "normal" high-luminosity galaxies with a moderate gas content (see, e.g., review [34]). Therefore, the question of the formation of external disks is of a more general nature.

Although the origin of massive outer gaseous disks in different VHR galaxies may be different, the large angular momentum of the gas apparently plays a key role in this process [10]. In the vicinity of VHR galaxies, separate HI regions are often observed, which can be associated with small satellites rich of gas or with diffuse clouds of neutral gas [37], but their masses are not comparable to the giant mass of gas in the extended disks. Accretion of gas from the environment (probably, from filaments) that took place in the past is more likely. It is in a good agreement both with the large specific angular momentum of gas and with the induction that, according to [38], galaxies with the highest mass of HI are located predominantly in filaments.

It is not clear whether the extended gaseous disks appeared at an early stage of formation of galactic disks, or they formed mainly in an epoch close to the modern one. Both options are possible, however, according to [8], the disks of VHR galaxies do not differ in pronounced asymmetry compared to the disks of "normal" galaxies in the control sample, and their centers coincide with the optical centers of galaxies even more accurately than in the control galaxies (of the Bluedisk sample). This argues against the recent formation of gaseous disks as a result of accretion or merging. This is also evidenced by the usually observed regular rotation of the outer gaseous disks.

As noted in the Introduction, the decrease in the HI density along the radius in late-type spiral galaxies, as a rule, has a similar character. In particular, it explains the close correlation between the integral mass of HI and the radius $R_{HI}$, which is tied to a certain boundary value of the surface density of HI. The VHR galaxies also lie (albeit with a significant scatter) along the general dependence $M_{HI}$–$R_{HI}$ constructed for "normal" galaxies [8, 10]. It indicates a nonrandom gas density distribution even outside the optical radius, which would hardly be expected in the case of a recent or ongoing accretion. An estimate of the stability parameter of gaseous disks of galaxies leads to the same conclusion (see below).

For all VHR galaxies considered here, the rotation curves are known, which makes it possible to estimate the degree of stability of the gaseous disks to gravitational perturbations. Figure 5 shows the distribution of the stability parameter $Q_{gas}$ for the gaseous component of the considered VHR galaxies. The role of the stellar disk is ignored here (its density decreases with distance faster than the gas density), the gaseous disk is assumed to be thin, and the gas velocity dispersion $c_g$ is taken as 10 km/s (the same as in the outer regions of optical disks). For three galaxies (PGC 33645 = UGC 6168, PGC 42836 = UGC 7899, PGC 47932 = NGC 5230), the dependences $Q_{gas}(R)$ were taken from [15], where they were obtained after taking into account the molecular gas in the inner region of the disks. Within the optical radius, the parameter $Q_{gas}$ for gas is in most cases above the critical value $Q_{gas}^c \approx 1.4$–$2$ for threshold stability. The stellar component of the disk, if it exists, can only bring the outer disk of the galaxy closer to a marginally stable state.

Among the galaxies, for which Fig. 5 shows the radial profiles $Q_{gas}$, only in two cases (PGC 21343 and PGC 63796) the stability parameter remains high ($Q_{gas} > 4$) at all $R$, which indicates a large reserve of gravitational stability. But at the same time, as noted above, the galaxy PGC 63796, which has the highest Toomre parameter, is observed edge-on, so the HI mass in its outer disk is apparently underestimated. For other galaxies, the minimum value of $Q_{gas}$ does not exceed 2.5–3. Thus, despite the low gas density in the extended HI disks, in most VHR galaxies there is a rather extended region of radial distances where the stability parameter $Q_{gas}$ is approximately constant. Although in all cases it indicates gravitational stability of gas disks, for most of the galaxies, it turns out to be close to the threshold value for stability.

Note that the value of $c_g \sim 10$ km/s accepted here at the far periphery of the disks may actually be overestimated. The measurements of the radial profile $c_g(R)$ obtained for about two dozen VHR galaxies of the Bluedisk sample [8] observed at the WSRT radio interferometer [14] showed, that in some galaxies at a distance of 20–40 kpc from the center $c_g$ decreases to 5–7 km/s. For such values of $c_g$, the outer disks of the VHR galaxies considered here would be even closer to the threshold stability.

The approximate constancy of $Q_{gas}$, which remains over a large radial extent, suggests that the gas density is closely related to the local angular velocity (epicyclic frequency) of the disks, on which the parameter $Q_{gas}$ depends. A similar conclusion was obtained for the gaseous disks of normal galaxies in the THINGS survey (see [39]). Recent or current accretion of gas onto the disk, which is random, does not explain the relationship between the gas density and the dynamic parameters of the disk, and therefore is unlikely to be a universal mechanism that determines the gas density distribution.

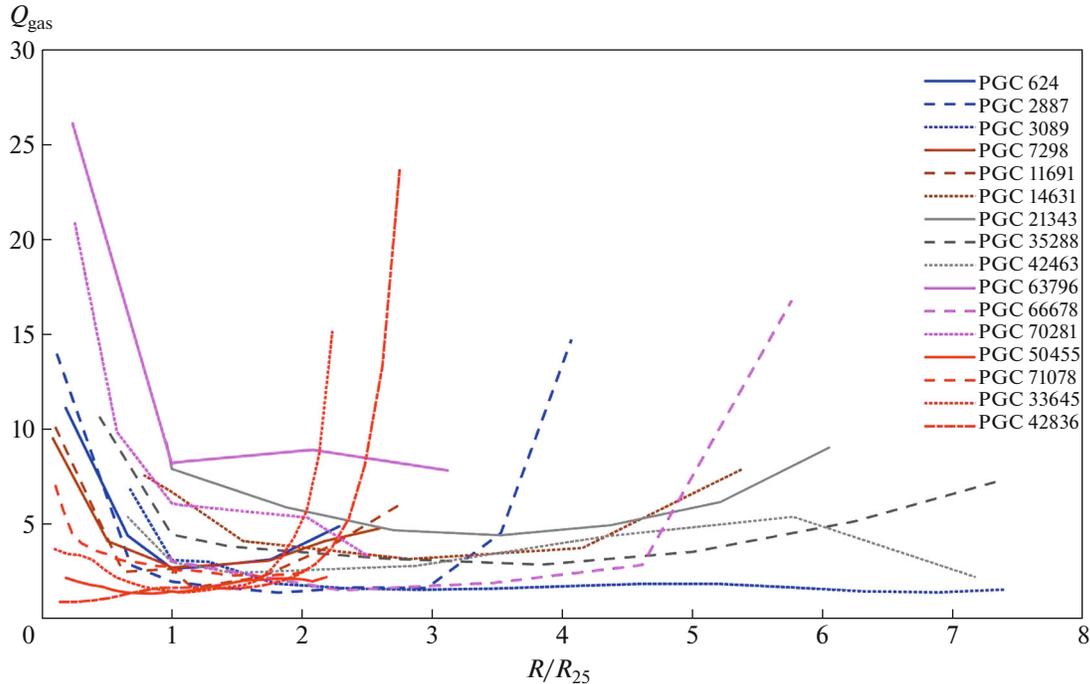

**Fig. 5.** Radial profiles of the Toomre parameter $Q_{\mathrm{gas}}$ for the considered VHR galaxies.

At a surface gas density not exceeding a few $M_\odot/\mathrm{pc}^2$, the wavelength of the most rapidly growing gravitational perturbations $\lambda_T = 2c_g^2/G\Sigma_g$ reaches a few kiloparsec, so this large-scale instability is not directly related to local star formation sites observed in the extended disks of some VHR galaxies. However, a gas density close to the threshold value can support spiral density waves and therefore affect star formation at smaller spatial scales. It is noteworthy that it is in those galaxies where the parameter $Q_{\mathrm{gas}}(R)$ falls below 2.5 or 2, i.e., gaseous disks are close to the state of marginal stability, a noticeable low-brightness spiral structure with star-forming regions inside of the spiral arms is observed in the outer disks. The most extended spirals are observed in PGC 3089 and PGC 676678, which can be traced to a distance of more than $3R_{25}$ (which is still less than $R_{\mathrm{HI}}$).

The efficiency of star formation decreases with decreasing gas density in both ordinary and VHR galaxies, and becomes very low at large distance from the center, so that its reciprocal value—the gas consumption time—should exceed the cosmological age of galaxies ($10^{10}$ years) there ([8, 15]). Therefore, the current rates of star formation are too low to have a significant effect on the distribution of gas in the outer gaseous disks of the considered galaxies on scales of several billion years. Then, the question arises why the outer gas-eous disks turned out to be close to the marginal stability over a large distance from the center.

It is natural to assume that the connection between the observed surface density of gas in the disk and the threshold stability condition was established in the past, billions of years ago, after the completion of the violent stage of formation of the stellar disk, when the gravitational instability of the gas layer played the role of a regulator of the star formation rate (this stage is observed now in galaxies with active star formation with a large redshift $z > 1$). After the inflow of gas into the galaxy became low, and as the result of star formation the density of the remaining gas decreased to a level excluding large-scale gravitational instability, the star formation rate and gas velocity dispersion so decreased (see, for example, the numerical model of the formation of the disk of the Galaxy proposed in [40]). As applied to ordinary spiral galaxies, this scenario was proposed in [3], but it can also be used for the extended disks of VHR galaxies. The main difference between these two cases is that in the extended disks of VHR galaxies, star formation almost completely ceased during the transition to the gravitationally stable state due to the very low gas density in the marginally stable disk, while in regions closer to the center of ordinary spiral galaxies, the efficiency of star formation remained at a higher level, and as the result the disks acquired a higher stability reserve over time. The radial distribution of the gas density in the outer regions of galaxies turned out for a long time to be associated with the $Q_{\mathrm{gas}}(R)$ profile close to critical one, which was established during the transition from the violent to the passive stage of evolution.

In such a scenario, some fraction of the gas in the extended disks of VHR galaxies first had to turn into stars in order for the surface density of the gaseous component of the disk to fall to the level of marginal stability. Therefore, it can be expected that the outer disks of VHR galaxies contain, in addition to gas, an "old" low-brightness stellar component that formed during intense star formation at the initial active stage of disk evolution. The expected surface density of the stellar population that could be formed at the active stage of disk formation should be at least comparable in magnitude to the observed surface density of gas at the periphery of galaxies ($\sim 1\,M_\odot/\text{pc}^2$ and below), otherwise star formation would have little effect on the resulting density of the gaseous disk. Taking the ratio $M/L_B \approx 3$ (in solar units) for the old metal-depleted stellar population (see the model [41]), we find that this corresponds to a surface brightness of about $28^m$/sq. arcsec, which can only be measured with very deep photometry.

Thus, stars that appeared at an early stage of the outer disk formation are expected to form an extended, very low-brightness old stars component. Its color should match the old and low metal stellar population.

Note that the idea that some part of the mass in the outer regions of galactic disks can remain undetected and belong to low-luminosity stars was previously mentioned in [42].

Nevertheless, in the outer disks of some VHR galaxies, a weak star formation is still observed, supported by low contrast spiral arms. In the absence of accretion or absorption of satellites, it should lead to a very slow decrease in the gas density.

A similar situation occurs for giant low surface brightness galaxies (GLSB galaxies) with a high gas content. The main amount of gas there is also distributed in an extended disk, and a spiral structure is often observed at large distances from the center. Observations show that the surface density of gas in the extended disks, at least in some GLSB galaxies, is also close to the critical value of the stability parameter. In [35], this conclusion was made for two of the four GLSB galaxies studied. According to the estimates obtained in [43] for several GLSB galaxies, the minimum values of the Toomre parameter for them are 2−3 units, which indicates the stability of their disks, but not a strong "overheating." The authors of [44] came to similar conclusions regarding the disk of the GLSB galaxy Malin-2. For 4 out of 7 GLSB galaxies considered in this paper, the observational data are in good agreement with the two-stage scenario of galaxy formation, when a galaxy of normal surface brightness is born, around which an LSB disk is then formed as a result of gas accretion from outside.

Summarizing, we emphasize the possibility of various reasons why a disk galaxy acquires or retains an anomalously high HI content in the disk. For some galaxies, where the gas is concentrated mainly in the inner regions of the disks, the role of the current gas accretion onto the disk cannot be ruled out (see Section 2). For example, for three galaxies in the High-Mass sample (UGC 7899, UGC 9037, and UGC 9334), the $H_\alpha$ velocity distribution showed the presence of non-circular gas motions, which indirectly confirm the possibility of a cold mode of gas accretion into the inner regions of galaxies [45]. At least in one case (the galaxy AGC 10111 with an excess of HI), the gaseous disk exhibits reverse rotation with respect to the stellar bulge, which definitely indicates the fall of gas onto the galaxy [13]. However, for VHR galaxies with massive extended outer HI disks rotating in the same direction as the stellar disk, two commonly proposed reasons for the high HI content (either accretion of gas with the same direction of the angular momentum vector, or the suppressed star formation, contributing to the accumulation and conservation of gas) can act together: first, the accretion, which has formed billions of years ago a gaseous disk with a large angular momentum, more dense than it is now, then a drop in gas density, and subsequent passive evolution with a very low efficiency of star formation. The approximate constancy of the stability parameter $Q_{\text{gas}}$ over a large range of radial distances in the VHR galaxies considered above suggests that their disks continue to "remember" the gas density distribution at which a transition from the violent stage of evolution to the passive one took place. The formation of disks of GLSB galaxies can also follow a similar scenario.

## FUNDING

This work was supported by RFBR grant 20-02-00080A.

*Translated by T. Sokolova*